\documentclass{article}%
\usepackage{amsmath}
\usepackage{graphicx}%
\usepackage{amsfonts}%
\usepackage{amssymb}
\def\be{\begin{equation}}
\def\ee{\end{equation}}
\def\bea{\begin{eqnarray*}}
\def\eea{\end{eqnarray*}}

\begin{document}

\title{Morphing quantum mechanics and fluid dynamics}
\author{{\Large Thomas Curtright}$\,^{\S}${\Large \ and David Fairlie}$\,^{\boxtimes
}\bigskip$\\$^{\S}\,$Department of Physics, University of Miami, \\Coral Gables, Florida 33124-8046, USA\\$^{\boxtimes}\,$Department of Mathematical Sciences, University of Durham, \\Durham, DH1 3LE, UK\\{\small Curtright@physics.miami.edu\ \ \ \ \ David.Fairlie@durham.ac.uk}}
\maketitle

\begin{abstract}
We investigate the effects of given pressure gradients on hydrodynamic flow
equations. \ We obtain results in terms of implicit solutions and also in the
framework of an extra-dimensional formalism involving the
diffusion/Schr\"{o}dinger equation.

\end{abstract}

\section{Introduction}

There has recently been renewed interest in the similarities between the
equations of fluid dynamics and those of quantum mechanics \cite{Polyakov}%
-\cite{Boldyrev}. \ This is a venerable subject \cite{Madelung}. \ In a
previous article \cite{CF} we have shown how any solution of the
multi-variable, multi-dimensional generalization of the simplest flow
equation, the Euler-Monge equation, is related to a particular type of
solution to a linear diffusion or heat equation, in twice the number of
``spatial'' dimensions. \ But, as is well-known, the heat equation when
complexified is just the Schr\"{o}dinger equation. \ So the method of
\cite{CF} may be viewed as a new transformation relating the Euler-Monge and
dimensionally-doubled Schr\"{o}dinger equations, which differs from the
transformation of Madelung.

Here, we extend this result, especially in the case of one space dimension, to
a flow equation driven by some particular, given pressure terms (i.e.
so-called ``body forces''). \ When the given pressure-gradient is either $t$
or $x$ independent, we obtain the general solution for such pressure driven
flows in closed but implicit form, and we relate these solutions to those of
driven diffusion/Schr\"{o}dinger equations. \ The corresponding
Schr\"{o}dinger equation contains a potential whose form is related to the
pressure-gradient. \ We also find explicit series solutions in a variety of
situations. \ Special attention is given to constant and linear
pressure-gradients. \ We discuss the complexification required to obtain the
Schr\"{o}dinger equation, and we emphasize analogies between the heat equation
and the Wigner-Moyal equation for the evolution of quantum mechanical
densities. \ In addition, we show how solutions of a related Bateman equation
may be constructed.

Should there be any lack of motivation to consider such matters, perhaps a
brief historical overview is warranted. \ The Euler-Monge (E-M) equations
embodied in fluid dynamics first appeared in 18th and 19th century studies of
both fluid flow \cite{Euler} and analytic geometry \cite{Monge}. \ Riemann
took up a study of the equations in the context of gas dynamics, discussing
the equations as a theory of invariants \cite{Riemann}\ (for a modern textbook
treatment, see \cite{Debnath}). \ His approach is widely applicable to almost
\emph{all} nonlinear flow problems, although it does not triumph over
turbulence. \ A systematic modern discussion of the E-M equations that
synthesizes ideas from both geometry and invariance theory can be found in the
review by Dubrovin and Novikov \cite{Dubrovin}. \ Most contemporary texts and
reviews stress the \emph{universal} role played by these nonlinear transport
equations in accordance with Whitham's theory \cite{Whitham}: \ Essentially
all nonlinear waves, even those in dispersive and dissipative media, involve
E-M equations, or simple variants of them, if the nonlinear wavetrains are
slowly varying. \ This makes the equations particularly useful for analyzing
the asymptotic behavior of nonlinear solutions. \ The E-M equations and their
conservation laws also serve as a useful starting point in Polyakov's study of
turbulence \cite{Polyakov} but without yet leading to a general solution of
the Navier-Stokes equations. \ In short, the Euler-Monge equations appear
widespread across a very broad landscape of physics and applied mathematics
problems, and therefore it is important to understand their solutions at as
many levels as possible. \ To that end we shall map some solutions of the E-M
equations with given pressure-gradients onto linear diffusion/Schr\"{o}dinger equations.

Suppose we ask for the condition that $\mathbb{U}(x,t,a)\equiv\frac{1}{a}\exp
au(x,t)$ is a solution of the following linear diffusion equation in two ($x$
and $a$) dimensions \cite{note1} with a given ``potential'' term $a\times
g(x,t)\,$:
\begin{equation}
\left(  \frac{\partial}{\partial t}-\frac{\partial^{2}}{\partial x\partial
a}-a\,g(x,t)\right)  \mathbb{U}(x,t,a)\,=0\;. \label{diffusion}%
\end{equation}
This holds if and only if a nonlinear equation is satisfied by $u=\frac{1}%
{a}\ln\left(  a\mathbb{U}\right)  $\thinspace:%
\begin{equation}
\frac{\partial u\left(  x,t\right)  }{\partial t}=u\left(  x,t\right)
\frac{\partial u\left(  x,t\right)  }{\partial x}+g(x,t)\;. \label{pressure}%
\end{equation}
We will call this the ``pressure driven Euler-Monge equation.'' \ \ The
function $g$ is the pressure-gradient, $g(x,t)=\partial_{x}p\left(
x,t\right)  $, as it would appear in the Navier-Stokes equation for an
incompressible fluid without viscosity. \ Here, we shall concentrate on
pressure-gradients which are \emph{given} functions, especially those which
are time-independent, $g(x)=\partial_{x}p\left(  x\right)  $, and not have $p$
determined by an equation of state. \ (Hence, our discussion is not truly a
description of ideal non-viscous fluid flow.)

The linear equation (\ref{diffusion}) can always be attacked, at least
formally, by the propagator method. \ However, in general, it is not possible
to find explicit, closed-form solutions, either to that linear equation or to
the associated nonlinear equation (\ref{pressure}). \ Nevertheless, general
implicit solutions to the nonlinear equation can be obtained in closed form,
as shown below, when the pressure-gradient is either a function of $x$ alone,
or of $t$ alone. \ Moreover, in cases where the pressure takes simple forms,
special explicit solutions exist, in terms of which the general solution may
be constructed.

\section{Simple examples}

In this section we discuss some simple cases before treating a general
pressure term. \ The technique we use is a variant of the well-known method of characteristics.

First, consider the case of the constant pressure equation, with $g\left(
x\right)  =0$.
\begin{equation}
\frac{\partial u}{\partial t}=u\frac{\partial u}{\partial x}\;.
\label{EulerMonge}%
\end{equation}
Two elementary special solutions are given by
\begin{equation}
x+ut=0\;,\;\;\;u=0\;. \label{ConstantPSolns}%
\end{equation}
Now, to obtain the general solution, just set one of these expressions to be
an arbitrary differentiable function of the other.
\begin{equation}
x+ut=G(u)\;, \label{Implicit}%
\end{equation}
for any differentiable $G$, where in (\ref{Implicit}), $u=u\left(  x,t\right)
$. \ This result is well-known \cite{Whitham}. \ By taking $x$ and $t$ partial
derivatives, the reader will easily verify that as a consequence of
(\ref{Implicit}), $u\left(  x,t\right)  $ must satisfy the Euler-Monge
equation (\ref{EulerMonge}).

Next, consider the equation resulting from a linear pressure, $p\left(
x\right)  =p\left(  0\right)  +kx$, hence constant pressure-gradient,
$g\left(  x\right)  =k$.
\begin{equation}
\frac{\partial u}{\partial t}=u\frac{\partial u}{\partial x}+k\;.
\label{EMConstantGrad}%
\end{equation}
Two special solutions are given by
\begin{equation}
x+ut-\frac{1}{2}kt^{2}=0\;,\;\;\;u-kt=0\;.
\end{equation}
Then the general solution is given by
\begin{equation}
x+ut-\frac{1}{2}kt^{2}=G(u-kt)\;, \label{LinearPSolns}%
\end{equation}
where again $G$ is any differentiable function, and where $u=u\left(
x,t\right)  $ in this last equation.

In the case of a quadratic pressure the equation becomes
\begin{equation}
\frac{\partial u}{\partial t}=u\frac{\partial u}{\partial x}+k^{2}x\;,
\label{EMLinearGrad}%
\end{equation}
with a linear forcing term $k^{2}x$. \ A particular solution is now given by
\begin{equation}
kx\cos(kt)+u\sin(kt)=0\;.
\end{equation}
(Dividing by $k$ and taking the limit $k\rightarrow0$ recovers the first of
(\ref{ConstantPSolns}).) \ Now we can construct another solution by
translating $kt$ by $\frac{\pi}{2}$ to give
\begin{equation}
kx\sin(kt)-u\cos(kt)\,=\,0\;.
\end{equation}
(In the $k\rightarrow0$ limit this gives the trivial solution in
(\ref{ConstantPSolns}).) \ The general solution is now given implicitly by
\begin{equation}
kx\cos(kt)+u\left(  x,t\right)  \sin(kt)=G(kx\sin(kt)-u\left(  x,t\right)
\cos(kt))\;,
\end{equation}
with $G$ an arbitrary differentiable function. \ 

\section{Recipe for a general implicit solution}

One way of proceeding with equation (\ref{pressure}), in the case of arbitrary
time-independent $g\left(  x\right)  ,$ is not to consider $u$ as a function
of $x$ and$\ t$, but to think of $t$ as a function of $x$ and$\ u$, i.e. the
hodograph method \cite{Whitham}. \ Then we have
\begin{equation}
\frac{\partial u}{\partial x}=-\,\frac{\partial t\left(  x,u\right)  /\partial
x}{\partial t\left(  x,u\right)  /\partial u}\;,\;\;\;\frac{\partial
u}{\partial t}=\frac{1}{\partial t\left(  x,u\right)  /\partial u}%
\end{equation}
This transforms equation (\ref{pressure}) into a \emph{linear} equation for
$t$.
\begin{equation}
\frac{\partial}{\partial x}\left(  u\,t\left(  x,u\right)  +x\right)
=g(x)\frac{\partial}{\partial u}t\left(  x,u\right)  \;. \label{ImplicitEqn}%
\end{equation}
The solution for general $g(x)$ is easily constructed from this. \ We turn
this equation into an integro-differential equation by integration with
respect to $x$.%

\begin{equation}
t\left(  x,u\right)  =F(u)-\frac{x}{u}+\frac{1}{u}\frac{\partial}{\partial
u}\int_{0}^{x}t\left(  z,u\right)  g(z)dz\;,
\end{equation}
where $F(u)$ is an arbitrary function of integration. \ This equation is then
formally solved by iteration:
\begin{gather}
t\left(  x,u\right)  =F\left(  u\right)  -\frac{x}{u}+\frac{1}{u}%
\frac{\partial}{\partial u}\int_{0}^{x}dz_{1}\,g(z_{1})\,\left(  F\left(
u\right)  -\frac{z_{1}}{u}\right)  +\frac{1}{u}\frac{\partial}{\partial u}%
\int_{0}^{x}dz_{1}\,g(z_{1})\,\frac{1}{u}\frac{\partial}{\partial u}\int
_{0}^{z_{1}}dz_{2}g(z_{2})\,t\left(  z_{2},u\right) \nonumber\\
=F\left(  u\right)  -\frac{x}{u}+\cdots+\int_{0}^{x}dz_{1}\,g(z_{1})\cdots
\int_{0}^{z_{n-1}}dz_{n}\,g(z_{n})\left(  \frac{1}{u}\frac{\partial}{\partial
u}\right)  ^{n}\left(  F\left(  u\right)  -\frac{z_{n}}{u}\right)  +\cdots\;.
\end{gather}
Since every term in the expansion is a product of a function of $x$ with a
function of $u$, all these terms are easily disentangled. \ The terms linear
in $F$ can be evaluated separately to give%
\begin{align}
&  \left(  1+\int_{0}^{x}dz_{1}\,g(z_{1})\,\frac{1}{u}\frac{\partial}{\partial
u}+\int_{0}^{x}dz_{1}\,g(z_{1})\int_{0}^{z_{1}}dz_{2}\,g(z_{2})\,\left(
\frac{1}{u}\frac{\partial}{\partial u}\right)  ^{2}+\cdots\right)  F\left(
u\right) \nonumber\\
&  =\sum_{n=0}^{\infty}\frac{1}{n!}\left(  \int_{0}^{x}dz\,g(z)\right)
^{n}\,\left(  \frac{1}{u}\frac{\partial}{\partial u}\right)  ^{n}F\left(
u\right)  =e^{\int_{0}^{x}dz\,g(z)\,\frac{1}{u}\frac{\partial}{\partial u}%
}F\left(  u\right)  =F\left(  u\sqrt{1+\frac{2}{u^{2}}\int_{0}^{x}%
dz\,g(z)}\right)  \;.
\end{align}
The $\left(  n+1\right)  $th term in the remaining series of terms is given by
$(-1)^{n+1}(2n-1)!!/u^{2n+1}$ times%

\begin{equation}
\int_{0}^{x}dz_{1}\,g(z_{1})\int_{0}^{z_{1}}dz_{2}\,g(z_{2})\cdots\int
_{0}^{z_{n-1}}dz_{n}\,g(z_{n})\,z_{n}=\frac{1}{\left(  n-1\right)  !}\int
_{0}^{x}\left(  p\left(  x\right)  -p\left(  z\right)  \right)  ^{n-1}%
\,g\left(  z\right)  \,z\,dz\;,
\end{equation}
where the pressure is $p\left(  x\right)  =p\left(  0\right)  +\int_{0}%
^{x}dz\,g(z)$. \ These terms can be summed. \ After an integration by parts,
the final result is \
\begin{equation}
t\left(  x,u\right)  =F\left(  u\sqrt{1+\frac{2}{u^{2}}\left(  p\left(
x\right)  -p\left(  0\right)  \right)  }\right)  -\frac{1}{u}\int_{0}%
^{x}\frac{dz}{\sqrt{1+\frac{2}{u^{2}}\left(  p\left(  x\right)  -p\left(
z\right)  \right)  }}\;. \label{ImplicitSolution}%
\end{equation}
This is the general solution to (\ref{ImplicitEqn}) with boundary value
$t\left(  x=0,u\right)  =F\left(  u\right)  $.

The result (\ref{ImplicitSolution}) is easily checked to be a solution to
(\ref{ImplicitEqn}). \ We recognize this as again being of the form where an
arbitrary function of one particular solution of (\ref{pressure}) is set equal
to a second particular solution. \ In this case $F\left(  u\sqrt
{1+\frac{2}{u^{2}}\left(  p\left(  x\right)  -p\left(  0\right)  \right)
}\right)  \equiv G\left(  \frac{1}{2}u^{2}+p\left(  x\right)  \right)  $
involves an arbitrary function of the combination that gives a particular
time-independent solution, $\frac{1}{2}u^{2}+p\left(  x\right)
=\mathrm{constant}$, of the time-independent special case of (\ref{pressure}),
$\frac{1}{2}\partial_{x}\left(  u^{2}\right)  =-\partial_{x}p\left(  x\right)
$, and the remaining terms in $t\left(  x,u\right)  $ implicitly furnish
another particular solution of the fully time-dependent (\ref{pressure}).

Similar statements apply to the situation where the driving term is a function
of only $t$, and not $x$. \ Thus $\partial u/\partial t=u\partial u/\partial
x+k(t)$ is solved implicitly by $x+ut-\int_{0}^{t}k\left(  \tau\right)  \tau
d\tau=G\left(  u-\int_{0}^{t}k\left(  \tau\right)  d\tau\right)  $, \ where
the latter function has as its argument the same form as the particular
$x$-independent solution $u-\int_{0}^{t}k\left(  \tau\right)  d\tau
=\mathrm{constant}$.

\section{Series solutions}

\paragraph{Building up to the extra dimension}

Of course, the existence of an implicit solution, while providing a pleasing
theoretical establishment of integrability, is not much use in many practical
applications. \ So we also look for power series expansions in $t$. \ Consider
first the case of a constant pressure-gradient,%
\begin{equation}
\frac{\partial u}{\partial t}=u\frac{\partial u}{\partial x}+k\;.
\end{equation}
Further time differentiation removes the explicit $k$, just as it would for
any time-independent $g\left(  x\right)  $,
\begin{equation}
\frac{\partial^{2}u}{\partial t^{2}}=\frac{1}{2}\frac{\partial}{\partial
t}\frac{\partial}{\partial x}u^{2}=\frac{\partial}{\partial t}(u\frac{\partial
u}{\partial x})=\frac{\partial}{\partial x}(u\frac{\partial u}{\partial t})\;.
\end{equation}
If the power series is denoted by
\begin{equation}
u(x,t)=\sum_{n=0}^{\infty}t^{n}u_{n}(x)\;,
\end{equation}
we have $u\left(  x,0\right)  \equiv u\left(  x\right)  =u_{0}$,
$\ u_{1}=k+\frac{1}{2}\frac{\partial}{\partial x}u\left(  x\right)  ^{2}$ ,
\ and so on. \ That is,%
\begin{equation}
(n+1)u_{n+1}=\frac{1}{2}\frac{\partial}{\partial x}\sum_{j=0}^{n}u_{j}%
u_{n-j}\;,\;\;\text{for}\;n\geq1\;.
\end{equation}
The resulting expression for $\left.  \partial^{n}u/\partial t^{n}\right|
_{t=0}=n!u_{n}$ for $n\geq2$ is therefore given, for constant $k,$ by
\begin{equation}
\left.  \frac{\partial^{n}u}{\partial t^{n}}\right|  _{t=0}=\sum
_{j=0}^{\left\lfloor n/2\right\rfloor }\left(  \frac{k}{2}\right)
^{j}\frac{n!}{j!\left(  n-2j\right)  !}\frac{\partial^{n-j-1}}{\partial
x^{n-j-1}}\left(  u\left(  x\right)  ^{n-2j}\frac{\partial u\left(  x\right)
}{\partial x}\right)  \;. \label{nthTimeDeriv}%
\end{equation}
Here, the floor function, $\left\lfloor n/2\right\rfloor $, is the greatest
integer less than or equal to $n/2.$ \ Hence,%
\begin{align}
u(x,t)  &  =u(x)+kt+\sum_{n=1}^{\infty}t^{n}\sum_{j=0}^{\left\lfloor
n/2\right\rfloor }\frac{1}{j!(n-2j+1)!}\left(  \frac{k}{2}\right)
^{j}\frac{\partial^{n-j}}{\partial x^{n-j}}u\left(  x\right)  ^{n-2j+1}%
\nonumber\\
&  =u(x)+kt+\sum_{n=1}^{\infty}t^{n}\sum_{j=0}^{\left\lfloor n/2\right\rfloor
}\frac{1}{j!(n-2j+1)!}\left(  \frac{k}{2}\frac{\partial}{\partial x}\right)
^{j}\left.  \left(  \frac{\partial^{2}}{\partial a\partial x}\right)
^{n-2j}e^{au\left(  x\right)  }u\left(  x\right)  \right|  _{a=0}\;.
\label{SeriesSolution}%
\end{align}
This last expression involving the exponential, evaluated at $a=0$ after all
differentiations, evokes the extra-dimensional approach. \ 

To follow through on that idea, recall that Hermite polynomials are given by
(\cite{Abram}, \textbf{22.3.10})
\begin{equation}
H_{n}\left(  z\right)  =n!\sum_{j=0}^{\left\lfloor n/2\right\rfloor }%
\frac{1}{j!\left(  n-2j\right)  !}\left(  -1\right)  ^{j}\left(  2z\right)
^{n-2j}=\left(  2z\right)  ^{n}n!\sum_{j=0}^{\left\lfloor n/2\right\rfloor
}\frac{1}{j!\left(  n-2j\right)  !}\left(  \frac{-1}{4z^{2}}\right)  ^{j}\;.
\end{equation}
So, taking $\partial/\partial x$ of both sides of (\ref{nthTimeDeriv}), we
find, formally,
\begin{gather}
\left.  \tfrac{\partial^{n}}{\partial t^{n}}\tfrac{\partial}{\partial
x}u\right|  _{t=0}=\sum_{j=0}^{\left\lfloor n/2\right\rfloor }\tfrac{\left(
k/2\right)  ^{j}}{\left(  j\right)  !\left(  n-2j\right)  !}\tfrac
{\partial^{n-j}}{\partial x^{n-j}}\left(  u^{n-2j}u_{x}\right)  =\left.
\sum_{j=0}^{\left\lfloor n/2\right\rfloor }\tfrac{1}{\left(  j\right)
!\left(  n-2j\right)  !}\left(  \tfrac{k}{2}\tfrac{\partial}{\partial
x}\right)  ^{j}\left(  \tfrac{\partial^{2}}{\partial x\partial a}\right)
^{n-2j}e^{au}u_{x}\right|  _{a=0}\nonumber\\
=\left(  \tfrac{1}{\sqrt{-1}}\sqrt{\tfrac{k}{2}\tfrac{\partial}{\partial x}%
}\right)  ^{n}\;\left.  \sum_{j=0}^{\left\lfloor n/2\right\rfloor }\tfrac
{1}{\left(  j\right)  !\left(  n-2j\right)  !}\left(  -1\right)  ^{j}\left(
\tfrac{\sqrt{-1}\frac{\partial^{2}}{\partial x\partial a}}{\sqrt{\frac{k}%
{2}\frac{\partial}{\partial x}}}\right)  ^{n-2j}e^{au}u_{x}\right|
_{a=0}\nonumber\\
=\left(  \tfrac{1}{\sqrt{-1}}\sqrt{\tfrac{k}{2}\tfrac{\partial}{\partial x}%
}\right)  ^{n}\tfrac{1}{n!}\left.  H_{n}\left(  \tfrac{\sqrt{-1}%
\frac{\partial^{2}}{\partial x\partial a}}{2\sqrt{\frac{k}{2}\frac{\partial
}{\partial x}}}\right)  e^{au}u_{x}\right|  _{a=0}\;.
\end{gather}
This result allows us to perform the sum, to obtain
\begin{align}
\tfrac{\partial}{\partial x}u\left(  x,t\right)   &  =\sum_{n=0}^{\infty}%
t^{n}\tfrac{\partial}{\partial x}u_{n}=\sum_{n=0}^{\infty}\left(  \tfrac
{t}{\sqrt{-1}}\sqrt{\tfrac{k}{2}\tfrac{\partial}{\partial x}}\right)
^{n}\tfrac{1}{n!}\left.  H_{n}\left(  \tfrac{\sqrt{-1}\frac{\partial^{2}%
}{\partial x\partial a}}{2\sqrt{\frac{k}{2}\frac{\partial}{\partial x}}%
}\right)  \;e^{au\left(  x\right)  }u_{x}\left(  x\right)  \right|
_{a=0}\nonumber\\
&  =\left.  \exp\left(  \tfrac{2t}{\sqrt{-1}}\sqrt{\tfrac{k}{2}\tfrac
{\partial}{\partial x}}\tfrac{\sqrt{-1}\frac{\partial^{2}}{\partial x\partial
a}}{2\sqrt{\frac{k}{2}\frac{\partial}{\partial x}}}-\left(  \tfrac{t}%
{\sqrt{-1}}\sqrt{\tfrac{k}{2}\tfrac{\partial}{\partial x}}\right)
^{2}\right)  \;e^{au\left(  x\right)  }u_{x}\left(  x\right)  \right|
_{a=0}\nonumber\\
&  =\left.  \exp\left(  t\tfrac{\partial^{2}}{\partial x\partial a}+\tfrac
{1}{2}kt^{2}\tfrac{\partial}{\partial x}\right)  \;e^{au\left(  x\right)
}u_{x}\left(  x\right)  \right|  _{a=0}\nonumber\\
&  =\left.  e^{t\frac{\partial^{2}}{\partial x\partial a}+\frac{1}{2}%
kt^{2}\frac{\partial}{\partial x}}\;\frac{\partial}{\partial x}%
\,\frac{e^{au\left(  x\right)  }-1}{a}\right|  _{a=0}\;,
\end{align}
through use of the generating function (\cite{Abram}, \textbf{22.9.17})
$e^{2zs-s^{2}}=\sum_{n=0}^{\infty}\frac{1}{n!}s^{n}H_{n}\left(  z\right)  .$
\ Thus we conclude%
\begin{equation}
u\left(  x,t\right)  =kt+\left.  e^{t\frac{\partial^{2}}{\partial x\partial
a}+\frac{1}{2}kt^{2}\frac{\partial}{\partial x}}\;\left(  \tfrac{e^{au\left(
x\right)  }-1}{a}\right)  \right|  _{a=0}=kt+\left.  e^{t\frac{\partial^{2}%
}{\partial x\partial a}}\;\tfrac{1}{a}\left(  e^{au\left(  x+\frac{1}{2}%
kt^{2}\right)  }-1\right)  \right|  _{a=0}\;. \label{USolution}%
\end{equation}
The RHS clearly involves just the time evolution of $\mathbb{U}$, evaluated at
the extra dimension boundary, $a=0$, but under the action of a slightly
modified extension of the time-dependent kernel in \cite{CF}, such that the
initial data has a supplementary shift $x\rightarrow x+\frac{1}{2}kt^{2}$. \ 

Comparing this to the undriven situation, we see the physical interpretation
is simply that of a constant external acceleration imposed uniformly on the
nonlinear self-interaction of the fluid. \ If we ``fall'' along with the
self-interacting fluid, constantly accelerating with it, the effects of the
external acceleration are removed. \ That is to say, from Galilean covariance,
if $u\left(  x,t\right)  $ is any solution of (\ref{EulerMonge}), then
$u\left(  x+\frac{1}{2}kt^{2},t\right)  +kt$ is a solution of
(\ref{EMConstantGrad}).

\paragraph{Working down from the extra dimension}

The previous implicit solutions suggest another way to obtain results in terms
of an extra-dimensional equation: \ Incorporate one of the particular
solutions into the exponential along with $u\left(  x,t\right)  $. \ For
example, define
\begin{equation}
\mathfrak{U}\left(  x,t,a\right)  =\tfrac{1}{a}\left(  e^{au\left(
x,t\right)  -akx\tan(kt)}-1\right)  \;. \label{LinearGradientU}%
\end{equation}
Then we have%
\begin{equation}
\left(  \tfrac{\partial}{\partial t}-\tfrac{\partial^{2}}{\partial a\partial
x}-k\tan kt\left(  1+x\tfrac{\partial}{\partial x}+a\tfrac{\partial}{\partial
a}\right)  \right)  \mathfrak{U}\left(  x,t,a\right)  =e^{au\left(
x,t\right)  -akx\tan(kt)}\left(  \partial_{t}u\left(  x,t\right)  -u\left(
x,t\right)  \partial_{x}u\left(  x,t\right)  -k^{2}x\right)  \;.
\end{equation}
This gives an equivalence between the Euler-Monge equation with linear
pressure-gradients, (\ref{EMLinearGrad}), and solutions, of the form given in
(\ref{LinearGradientU}), to the following modified extra-dimensional linear
equation.
\begin{equation}
\left(  \frac{\partial}{\partial t}-\frac{\partial^{2}}{\partial a\partial
x}\right)  \mathfrak{U}\left(  x,t,a\right)  =\left(  k\tan kt\right)  \left(
1+x\frac{\partial}{\partial x}+a\frac{\partial}{\partial a}\right)
\mathfrak{U}\left(  x,t,a\right)  \;. \label{WEquation}%
\end{equation}
The formal solution to this linear equation is given by the propagator method.%

\begin{equation}
\mathfrak{U}\left(  x,t,a\right)  =\mathbf{K}\left(  t\right)  \;\mathfrak{U}%
\left(  x,t=0,a\right)  =\mathbf{K}\left(  t\right)  \;\tfrac{1}{a}\left(
e^{au\left(  x\right)  }-1\right)  \;,
\end{equation}
where the operator kernel obeys
\begin{equation}
\frac{\partial}{\partial t}\mathbf{K}\left(  t\right)  =\left(  \frac{\partial
^{2}}{\partial x\partial a}+\left(  k\tan kt\right)  \left(  1+x\frac{\partial
}{\partial x}+a\frac{\partial}{\partial a}\right)  \right)  \mathbf{K}\left(
t\right)  \;.
\end{equation}
That is, the kernel is a time-ordered exponential \cite{DeFacio} given by the
usual series of nested integrals.
\begin{gather}
\mathbf{K}\left(  t\right)  =1+\sum_{j=1}^{\infty}\int_{0}^{t}d\tau_{1}%
\int_{0}^{\tau_{1}}d\tau_{2}\cdots\int_{0}^{\tau_{j-1}}d\tau_{j}\left(
\tfrac{\partial^{2}}{\partial x\partial a}+k\tan k\tau_{1}\left(
1+x\frac{\partial}{\partial x}+a\frac{\partial}{\partial a}\right)  \right)
\times\nonumber\\
\times\left(  \tfrac{\partial^{2}}{\partial x\partial a}+k\tan k\tau
_{2}\left(  1+x\frac{\partial}{\partial x}+a\frac{\partial}{\partial
a}\right)  \right)  \cdots\left(  \tfrac{\partial^{2}}{\partial x\partial
a}+k\tan k\tau_{j}\left(  1+x\frac{\partial}{\partial x}+a\frac{\partial
}{\partial a}\right)  \right)  \;.
\end{gather}
This time-ordered exponential can be worked out as products of ordinary
exponentials, through use of the underlying algebra.%
\begin{equation}
\left[  A,C\right]  =0=\left[  B,C\right]  \;,\;\;\;\left[  A,B\right]  =A\;,
\label{algebra}%
\end{equation}
where $A=\frac{\partial^{2}}{\partial x\partial a}\;,\;B=a\frac{\partial
}{\partial a}\;,\;C=1+x\frac{\partial}{\partial x}-a\frac{\partial}{\partial
a}$. \ Note that $C$ is central. \ Writing%
\begin{align}
\frac{\partial^{2}}{\partial x\partial a}+\left(  k\tan kt\right)  \left(
1+x\frac{\partial}{\partial x}+a\frac{\partial}{\partial a}\right)   &
=\frac{\partial^{2}}{\partial x\partial a}+2\left(  k\tan kt\right)  \left(
a\frac{\partial}{\partial a}\right)  +\left(  k\tan kt\right)  \left(
1+x\frac{\partial}{\partial x}-a\frac{\partial}{\partial a}\right) \nonumber\\
&  =A+2\left(  k\tan kt\right)  B+\left(  k\tan kt\right)  C\;,
\end{align}
we see that we need to solve an evolution equation of the form%
\begin{equation}
\frac{d}{dt}\mathbf{K}\left(  t\right)  =\left(  \alpha\left(  t\right)
A+\beta\left(  t\right)  B+\gamma\left(  t\right)  C\right)  \mathbf{K}\left(
t\right)  \;,
\end{equation}
with $\mathbf{K}\left(  t=0\right)  =1$, where $A,B,C$ obey (\ref{algebra}),
and where the coefficients $\alpha,\beta,\gamma$ commute with everything.
\ The solution is given by the Baker-Campbell-Hausdorff technique.%

\begin{align}
\mathbf{K}\left(  t\right)   &  =\exp\left(  B\int_{0}^{t}\beta\left(
\tau\right)  d\tau\right)  \exp\left(  A\int_{0}^{t}\alpha\left(  \tau\right)
e^{\int_{0}^{\tau}\beta\left(  \tau^{\prime}\right)  d\tau^{\prime}}%
d\tau\right)  \exp\left(  C\int_{0}^{t}\gamma\left(  \tau\right)  d\tau\right)
\nonumber\\
&  =\exp\left(  A\int_{0}^{t}\alpha\left(  \tau\right)  e^{-\int_{\tau}%
^{t}\beta\left(  \tau^{\prime}\right)  d\tau^{\prime}}d\tau\right)
\exp\left(  B\int_{0}^{t}\beta\left(  \tau\right)  d\tau\right)  \exp\left(
C\int_{0}^{t}\gamma\left(  \tau\right)  d\tau\right)  \;.
\end{align}
For the case at hand, $\alpha\left(  t\right)  =1$, $\beta\left(  t\right)
=2k\tan kt=2\gamma\left(  t\right)  $, $\int_{0}^{t}\beta\left(  \tau\right)
d\tau=-\ln\cos^{2}kt=2\int_{0}^{t}\gamma\left(  \tau\right)  d\tau$, and

\noindent$\int_{0}^{t}\alpha\left(  \tau\right)  e^{-\int_{\tau}^{t}%
\beta\left(  \tau^{\prime}\right)  d\tau^{\prime}}d\tau=\frac{1}{2k}\sin2kt$.
\ So%
\begin{gather}
\mathbf{K}\left(  t\right)  =\exp\left(  \frac{1}{2k}A\sin\left(  2kt\right)
\right)  \exp\left(  -B\ln\cos^{2}kt\right)  \exp\left(  -\frac{1}{2}C\ln
\cos^{2}kt\right) \nonumber\\
=\frac{1}{\cos kt}\exp\left(  \frac{1}{2k}\sin\left(  2kt\right)
\frac{\partial^{2}}{\partial x\partial a}\right)  \exp\left(  -\frac{1}{2}%
\ln\left(  \cos^{2}kt\right)  \left(  x\frac{\partial}{\partial x}%
+a\frac{\partial}{\partial a}\right)  \right)  \;.
\end{gather}
The right-most exponential operator rescales $x$ and $a$ in the initial data,
and the left-most operator then evolves these data as in the undriven case,
except for the replacement $t\rightarrow\frac{1}{2k}\sin\left(  2kt\right)  $. \ 

Thus the action of this kernel on the initial, exponentiated data becomes%
\begin{equation}
\mathfrak{U}\left(  x,t,a\right)  =\mathbf{K}\left(  t\right)  \left(
\frac{e^{au\left(  x\right)  }-1}{a}\right)  =e^{\frac{1}{2k}\sin\left(
2kt\right)  \frac{\partial^{2}}{\partial x\partial a}}\;\left(
\frac{e^{u\left(  x/\cos kt\right)  \;a/\cos kt}-1}{a}\right)  \;.
\label{FrakUSolution}%
\end{equation}
As in \cite{CF}, series expansion in powers of $a$ and evaluation of
(\ref{FrakUSolution}) at $a=0$ produces a time series solution for $u\left(
x,t\right)  $, since $\lim_{a\rightarrow0}\mathfrak{U}\left(  x,t,a\right)
=\lim_{a\rightarrow0}\frac{1}{a}\left(  e^{au\left(  x,t\right)  -akx\tan
(kt)}-1\right)  =u\left(  x,t\right)  -kx\tan(kt)$. \ The series so obtained
explicitly exhibits any difference between $u$ and the particular solution
incorporated into the exponential.%
\begin{equation}
u\left(  x,t\right)  =kx\tan(kt)+\frac{1}{\cos kt}\sum\limits_{j=0}^{\infty
}\frac{1}{\left(  1+j\right)  !}\left(  \frac{1}{k}\sin kt\right)
^{j}\frac{d^{j}}{dx^{j}}\left(  u\left(  \tfrac{x}{\cos kt}\right)  \right)
^{1+j}\;. \label{LinearSeriesSolution}%
\end{equation}
Once having found these results so systematically -- although perhaps somewhat
laboriously -- their interpretation is almost as simple as for the case of a
constant driving term. \ As the reader may check by direct substitution, if
$u\left(  x,t\right)  $ is any solution of (\ref{EulerMonge}), then
$\frac{1}{\cos kt}\,u\left(  \frac{x}{\cos kt},\frac{1}{k}\tan kt\right)
+kx\tan kt$ is a solution of (\ref{EMLinearGrad}) resulting from the same
initial value data \cite{note2,note3,note4}.

The preceding modification of the propagator approach to (\ref{diffusion}) can
be employed for general pressure-gradients. \ If $u_{p}\left(  x,t\right)  $
is a particular solution to (\ref{pressure}), even if the pressure-gradient
depends on both $x$ \emph{and} $t$, then
\begin{equation}
\mathfrak{U}\left(  x,t,a\right)  =\tfrac{1}{a}\left(  e^{au\left(
x,t\right)  -au_{p}\left(  x,t\right)  }-1\right)
\end{equation}
satisfies the linear equation%
\begin{equation}
\left(  \frac{\partial}{\partial t}-\frac{\partial^{2}}{\partial a\partial
x}-\left(  \partial_{x}u_{p}\left(  x,t\right)  \right)  \left(
1+a\frac{\partial}{\partial a}\right)  -u_{p}\left(  x,t\right)
\frac{\partial}{\partial x}\right)  \mathfrak{U}\left(  x,t,a\right)  =0
\label{GeneralLinearEquation}%
\end{equation}
if and only if $\partial u\left(  x,t\right)  /\partial t=u\left(  x,t\right)
\partial u\left(  x,t\right)  /\partial x+g\left(  x,t\right)  $. \ Thus,
evolution of any exponentiated initial data, expressed as $\exp\left(
u\left(  x\right)  -u_{p}\left(  x\right)  \right)  $, can be effected by
solving a linear equation for the kernel appropriate to
(\ref{GeneralLinearEquation}), and acting with that kernel on $\mathfrak{U}%
\left(  x,t=0,a\right)  .$ \ For certain problems, in addition to the
$g\left(  x\right)  =k^{2}x$ case just discussed, this approach may be simpler
to carry through than a direct attempt to construct the propagator for
(\ref{diffusion}).

\paragraph{Breaking waves, shocks, and all that}

The formation of shocks under Euler-Monge evolution is well-known, and has a
simple qualitative description in terms of characteristics (for example, see
\cite{Debnath,Whitham}). \ It is perhaps not so well-known that the series
solution (\ref{SeriesSolution}) leads to the same conclusions. \ In general,
this may require some asymptotic approximations, but it is easy to find simple
analytical examples. \ 

A particularly elementary illustration of a breaking kinematic wave is given
by taking continuous, piecewise linear $u$ profiles. \ First consider the case
of the undriven Euler-Monge equation. \ The time power series sum
(\ref{SeriesSolution}) for $k=0$ is easily evaluated in this case. \ Take one
particular initial linear segment to be%
\begin{equation}
u\left(  x\right)  =\alpha+\beta x\;.
\end{equation}
Then we have $\frac{1}{(n+1)!}\frac{\partial^{n}}{\partial x^{n}}u\left(
x\right)  ^{n+1}=\beta^{n}\left(  \alpha+\beta x\right)  $, and so, for the
time development of this segment,%
\begin{equation}
u(x,t)=\left(  \alpha+\beta x\right)  \left(  1+\sum_{n=1}^{\infty}\left(
\beta t\right)  ^{n}\right)  =\frac{\alpha+\beta x}{1-\beta t}\;.
\end{equation}
The segment becomes vertical at its ``break-time'' $t_{\text{break}}=1/\beta$,
and while obvious, it is perhaps worth stressing that this is precisely the
same as the radius of convergence of the time power series
(\ref{SeriesSolution}).

Continuously rising and falling combinations of such linear segments, forming
a triangular ``bump,'' with equal amounts of rise and fall, have conserved
height and base length. \ Hence the triangle's area is constant. \ So too for
the area under a continuous polygonal bump. \ This area conservation is just
what you would expect from the Euler-Monge equation, written as a local
conservation law: $\ \partial_{t}u\left(  x,t\right)  =\partial_{x}\left(
u^{2}\left(  x,t\right)  /2\right)  $. \ At least, this is what you would
expect for continuous and \emph{differentiable} functions. \ The fact that the
same integrated conservation law also applies to continuous, piecewise
differentiable functions may be surprising upon first encounter, but it is
well-known to be true \cite{Debnath,Whitham}.

After the earliest such break-time, the continuous piecewise linear $u$
profile will no longer be single-valued. \ If there are physical reasons to
prohibit this, a standard procedure is to replace the continuous but
multi-valued $u$ with a shock discontinuity, positioned so as to preserve the
previously mentioned area conservation (cf. the Rankine-Hugoniot rule, or the
Lax entropy condition \cite{Lax}).

When constant pressure-gradients drive the Euler-Monge equation, the result
(\ref{SeriesSolution}) shows the same initial linear segment evolves as%
\begin{equation}
u(x,t)=kt+\frac{\alpha+\beta\left(  x+\frac{1}{2}kt^{2}\right)  }{1-\beta
t}\;,
\end{equation}
with the same break-time as before, $t_{\text{break}}=1/\beta$. \ When linear
pressure-gradients are added to the Euler-Monge equation,
(\ref{LinearSeriesSolution}) shows the initial linear segment evolves as%
\begin{equation}
u(x,t)=kx\tan kt+\frac{\alpha+\beta x/\cos kt}{\cos kt-\frac{\beta}{k}\sin
kt}\;,
\end{equation}
with modified break-time given by $t_{\text{break}}=\frac{1}{k}\arctan\left(
k/\beta\right)  $. \ Again, these break-times are just the radii of
convergence for the respective time power series.

Another example is given by taking exponential initial data, for which we take
the $k=0$ case to again see the wave evolution from the series
(\ref{SeriesSolution}). \ Suppose that initially $u\left(  x\right)
=Ae^{x/L}$, then $\frac{\partial^{n}}{\partial x^{n}}u\left(  x\right)
^{n+1}=\left(  \frac{n+1}{L}\right)  ^{n}\left(  Ae^{x/L}\right)  ^{n+1}$, so%
\begin{equation}
u(x,t)=Ae^{x/L}\left(  1+\sum_{n=1}^{\infty}\frac{\left(  n+1\right)  ^{n}%
}{(n+1)!}\left(  \frac{At}{L}e^{x/L}\right)  ^{n}\right)  =\frac{L}{t}%
\sum_{n=1}^{\infty}\frac{n^{n-1}}{n!}\left(  \frac{At}{L}e^{x/L}\right)
^{n}\;. \label{LambertSeries}%
\end{equation}
This series is familiar to anyone who has ever attempted to find by iteration
the inverse function for the elementary equation $x\left(  y\right)  =ye^{y}$,
i.e. to obtain $y\left(  x\right)  =xe^{-y}=x-x^{2}+\frac{3}{2}x^{3}%
-\frac{8}{3}x^{4}+-\;\cdots=x\sum_{n=1}^{\infty}\left(  -nx\right)  ^{n-1}%
/n!$. \ Lambert and Euler studied the analytic properties of a generalization.
\ There is a contemporary movement to standardize the function and name it
after Lambert \cite{Lambert}. \ The resulting ``Lambert W function'' series
definition, its log derivative, and its key functional relation are given by%
\begin{equation}
W\left(  z\right)  =-\sum_{n=1}^{\infty}\frac{1}{n!}n^{n-1}\left(  -z\right)
^{n}\;,\;\;\;\text{\ }z\frac{d}{dz}W\left(  z\right)  =\frac{W\left(
z\right)  }{1+W\left(  z\right)  }\;,\;\;\;W\left(  z\right)  \,\exp\left(
W\left(  z\right)  \right)  =z\;. \label{LambertProps}%
\end{equation}
This series converges absolutely for $\left|  z\right|  <1/e$.

So the above solution (\ref{LambertSeries}) for exponential initial data is
\begin{equation}
u(x,t)=-\,\frac{L}{t}\,W\left(  -\,\frac{At}{L}\,e^{x/L}\right)  \;.
\label{LambertFront}%
\end{equation}
The radius of convergence of the series for $u$ is therefore both $t$- and $x
$-dependent. \ Absolute convergence holds when $\frac{At}{L}e^{x/L}<1/e\;$.
\ Actually, the Lambert function is well-behaved for all positive arguments,
as is obvious from the inverse function, hence the exponential shape evolves
backward in time without the formation of any singularities. \ But for forward
time evolution, the wave always breaks, or forms a shock-front, and it does so
at earlier times for larger $x$. \ 

The condition for the wave to break, for positive $t$, is just that the radius
of convergence of the time power series is reached. \ That is $t_{\text{break}%
}=\frac{L}{A}e^{-1-x/L}$. \ At this time, for the given position, the slope of
the wave profile becomes vertical, since%
\begin{equation}
L\frac{d}{dx}u\left(  x,t\right)  =\frac{u\left(  x,t\right)  }{1-t\,u\left(
x,t\right)  /L}\;,
\end{equation}
and $W\left(  -1/e\right)  =-1$ gives $u\left(  x,t=\frac{L}{A}e^{-1-x/L}%
\right)  =L/t$. \ The motion of the vertical face of the wave is simply given
by $\frac{At}{L}e^{x/L}=\frac{1}{e}$ or%
\begin{equation}
x\left(  t\right)  =L\left(  \ln\frac{L}{A}-1-\ln t\right)  \;.
\label{FaceMotion}%
\end{equation}
There is also the implicit definition of $u\left(  x,t\right)  $ provided by
(\ref{Implicit}) when specialized to the case at hand: $\ x+ut=G(u)=L\ln
\left(  u/A\right)  $, or
\begin{equation}
u\left(  x,t\right)  =Ae^{x/L}\exp\left(  t\,u\left(  x,t\right)  /L\right)
\;.
\end{equation}
This functional relation is immediately seen to be the familiar one obeyed by
the Lambert function, as in (\ref{LambertProps}) above. \ Indeed, our
derivation of the series solution to the Euler-Monge equation for general data
obeying (\ref{Implicit}) may be thought of as a generalization of such series
solutions to an infinite class of such functional relations.

We plot below the $A=1=L,$ $k=0$ case of $u$ for various $t$. \ The
exponential initial data $u(x,0)=e^{x}$ is shown as a black curve. \ With our
sign conventions in the Euler-Monge equations, the wave moves to the left.
\ For negative $t$ the solution is also unbounded as $x\rightarrow+\infty$,
but non-singular for finite $x$, as shown by green curves in the right-most
part of the Figure for various negative times. \ For positive $t$, vertical
profiles occur where the curve has infinite slope, at the points indicated by
small circles on the tips of the series-defined segments of $u\left(
x,t\right)  $ shown by green curves in the left-most part of the Figure for
various positive times. \ The series fails to converge for larger values of
$x$, beyond these tips. \ (In more realistic situations, this may be
indicative of some missing physics, such as would be represented by adding
higher $x$ derivatives in the original PDE, as in Burgers' equation, see
\cite{Boldyrev}.) \ However, the implicit definition of $u\left(  x,t\right)
$ given by the functional relation (\ref{Implicit}) serves to define upper
branches of $u\left(  x,t\right)  $ that are shown by the orange curves
extending beyond the location of the vertical face. \ These upper branches may
be thought of as the overhanging portions of the breaking wave. \ A pretty
numerical example of a continuous, integrable, piecewise exponential wave is
obtained by joining rising and falling exponentials, so that the initial data
is $\exp\left(  -\left|  x\right|  \right)  $. \ The time evolution of this
example can be found graphically displayed on the website of one of the
authors. \ (\emph{http://curtright.com/waves.html})%
\begin{center}
\includegraphics[
height=3.6461in,
width=4.8577in
]%
{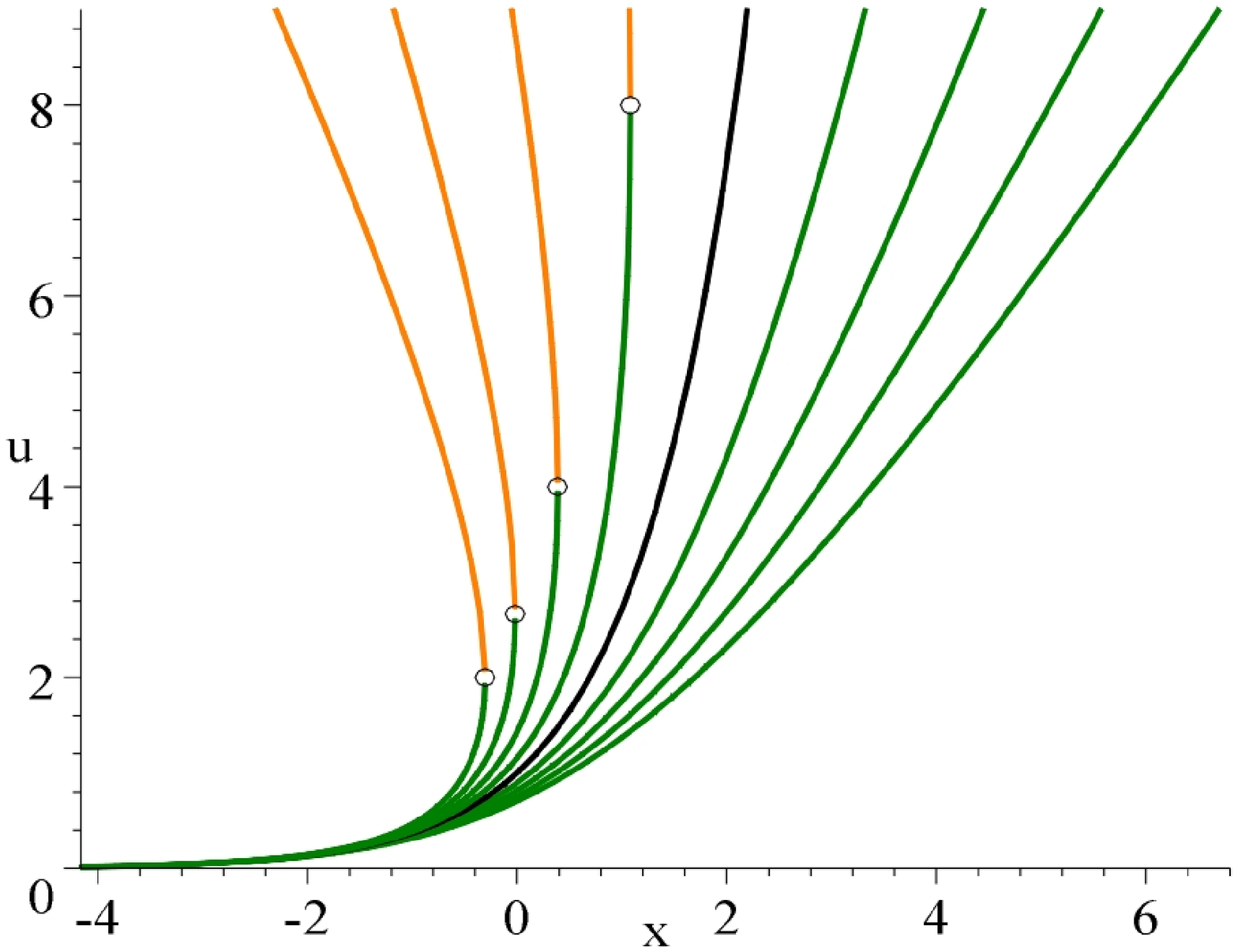}%
\\
{\small Evolution of the ``Lambert front''\ for }$t=-1/2$ to $+1/2$%
{\small \ in }$\Delta t=1/8 ${\small \ steps}%
\end{center}
\noindent The location of the vertical face moves to the left as $x\left(
t\right)  =-1-\ln t$, more slowly than the upper branch of the front, and
reaches the origin at $t=1/e$. \ (Recall the model is non-relativistic.)

Moreover, we may extend the solution (\ref{LambertFront}) of the Euler-Monge
equation stemming from the exponential initial data, using the analysis of the
preceding subsections, to find the constant body-force solution arising from
the same initial data.%
\begin{equation}
u(x,t)=kt-\frac{L}{t}\,W\left(  -\,\frac{At}{L}\exp\frac{1}{L}\left(
x+\frac{1}{2}kt^{2}\right)  \right)  \;.
\end{equation}
We also find the corresponding linear body-force solution.%
\begin{equation}
u(x,t)=kx\tan kt-\frac{Lk}{\sin kt}\,W\left(  -\,\frac{A\tan kt}{Lk}%
\exp\frac{x}{L\cos kt}\right)  \;.
\end{equation}
Thus we have conditions for the wave profile to be vertical (signaling the
local formation of shocks) whenever the arguments of the Lambert functions are
equal to $-1/e$, i.e. \
\begin{equation}
t_{\text{break}}=\frac{L}{Ae}e^{-x/L}\;,\;\;\;t_{\text{break}}=\frac{L}%
{Ae}\exp\frac{-1}{L}\left(  x+\frac{1}{2}kt_{\text{break}}^{2}\right)
\;,\;\;\text{or}\;\;\tan kt_{\text{break}}=\frac{Lk}{Ae}\exp\left(
\frac{-x}{L\cos kt_{\text{break}}}\right)  \;,
\end{equation}
respectively, for the three situations considered. \ The latter two conditions
are very implicit determinations of $t_{\text{break}}$. \ 

\section{Simple consequences of the extra dimension approach}

In this section a trivial method is given to construct an infinite number of
solutions to the diffusion equation starting from a solution of the Monge
equation, which might well be taken as one of the particular solutions of
explicit form. \ 

This follows very simply from the existence of operators which commute with
the diffusion operator. \ Consider the case where $p=0$, i.e the free case.
Then because $\frac{\partial}{\partial t},\ \frac{\partial}{\partial
a},\ \frac{\partial}{\partial x}$ and the ``boost'' generator $x\frac{\partial
}{\partial x}-a\frac{\partial}{\partial a}$ all commute with the operator
$\frac{\partial}{\partial t}-\frac{\partial^{2}}{\partial x\partial a}$, we
can find an infinite class of solutions of the diffusion equation based upon
the solution $u$ of the Monge equation by operating upon the ``primitive''
solution $\mathbb{U}(x,t,a)$ with arbitrary functions of these operators.
\ (In fact only three such operators are required, as the diffusion operator
is itself a combination of the others.) For example, consider
\begin{equation}
G(\frac{\partial}{\partial a})\frac{\partial}{\partial x}\mathbb{U}%
(x,t,a)=G(u)\frac{\partial u}{\partial x}e^{au(x,t)}\;,
\end{equation}
where $G$ is an arbitrary function with no singularity at the origin, or
\begin{equation}
G(\frac{\partial}{\partial a})\frac{\partial}{\partial t}\mathbb{U}%
(x,t,a)=G(u)\frac{\partial u}{\partial t}e^{au(x,t)}\;.
\end{equation}
Note that in fact this latter class is really the same as the former, on
account of the differential equation satisfied by $u$: All time derivatives
may be replaced by derivatives with respect to $x$. \ In the case of generic
$g(x)\neq0$, the only operator which commutes with the diffusion operator
(\ref{diffusion}) is $\frac{\partial}{\partial t}$. \ But there are
exceptional cases. \ For example, when $g(x)=k^{2}x$, the boost generator
$x\frac{\partial}{\partial x}-a\frac{\partial}{\partial a}$ also commutes.

Suppose $u$ is a particular solution of the Monge equation, in the case of the
linear pressure-gradient term, where $g(x)=k^{2}x$, such as one of the
solutions given above: $u=kx\tan(kt)$ or $u=-kx\cot(kt)$. \ Suppose further
that another solution of the diffusion equation is sought of the form
$\mathbb{U}(x,t,a)=v(x,t)\,e^{au(x,t)}$. \ Then it may be verified that
$v(x,t)$ satisfies the equation
\begin{equation}
\frac{\partial v}{\partial t}-\frac{\partial(uv)}{\partial x}=0\;.
\end{equation}
If $u$ is known explicitly then this equation can be integrated: \ In the case
$u=kx\tan(kt)$, $v=-1/\cos(kt)$, and in the case $u=-kx\cot(kt),$
$v=1/\sin(kt)$.

\section{Schr\"{o}dinger and Euler-Monge relations}

The diffusion equation may be interpreted as a Schr\"{o}dinger equation, upon
complexification. \ To be explicit, consider the time-dependent
Schr\"{o}dinger equation for a harmonic potential in one dimension.
\begin{equation}
\frac{\partial\psi(x,t)}{\partial t}=i\frac{\partial^{2}\psi(x,t)}{\partial
x^{2}}-\frac{1}{4}ik^{2}x^{2}\psi(x,t)\;.
\end{equation}
We can rewrite this without change of content, together with its complex
conjugate (with $(x,\ a)$ both real) as
\begin{align}
\frac{\partial\psi(x+ia,t)}{\partial t}  &  =\frac{\partial^{2}}{\partial
x\partial a}\psi(x+ia,t)-\frac{1}{4}ik^{2}(x+ia)^{2}\psi(x+ia,t)\;,\nonumber\\
\frac{\partial\bar{\psi}(x-ia,t)}{\partial t}  &  =\frac{\partial^{2}%
}{\partial x\partial a}\bar{\psi}(x-ia,t)+\frac{1}{4}ik^{2}(x-ia)^{2}\bar
{\psi}(x-ia,t)\;.
\end{align}
Multiply the first by $\bar{\psi}(x-ia,t)$ and the second by $\psi(x+ia,t)$
and add to obtain
\begin{equation}
\frac{\partial}{\partial t}\left(  \psi(x+ia,t)\bar{\psi}(x-ia,t)\right)
=\frac{\partial^{2}}{\partial x\partial a}\left(  \psi(x+ia,t)\bar{\psi
}(x-ia,t)\right)  +k^{2}ax\left(  \psi(x+ia,t)\bar{\psi}(x-ia,t)\right)  \;.
\end{equation}
Defining the \emph{complex} point-split density $\rho(x,t,a)=\psi
(x+ia,t)\bar{\psi}(x-ia,t)$, we see that $\rho$ obeys a diffusion type
equation which is solvable in terms of a one-dimensional Euler-Monge equation
with a linear pressure-gradient term, in the manner indicated above.
\ Defining as previously $x=\frac{1}{2}(x_{1}+ix_{2}),\ a=\frac{1}{2}%
(x_{1}-ix_{2})$, we find that $\rho$ satisfies formally a two-dimensional
diffusion equation with a complex harmonic potential:
\begin{equation}
\frac{\partial\rho}{\partial t}=\frac{\partial^{2}\rho}{\partial x_{1}^{2}%
}+\frac{\partial^{2}\rho}{\partial{x_{2}^{2}}}-\frac{1}{4}ik^{2}(x_{1}%
^{2}+x_{2}^{2})\rho\;.
\end{equation}
This transformation of the Schr\"{o}dinger equation is subtle, however, as now
the arguments of the original wave functions are complex for real $x_{1}$ and
$x_{2}$: \ $x\pm ia=\frac{1}{2}\left(  1\pm i\right)  \left(  x_{1}\pm
x_{2}\right)  $.

For \emph{real} point-splitting, there are an infinite number of conserved
charges and corresponding current densities for the free-particle
Schr\"{o}dinger equation, in an arbitrary number of spatial dimensions, as is
well-known. These conservation laws follow immediately from%
\begin{equation}
\partial_{t}\psi\left(  \mathbf{x},t\right)  =i\kappa\nabla^{2}\psi\left(
\mathbf{x},t\right)  \;,\;\;\;\partial_{t}\bar{\psi}\left(  \mathbf{x}%
,t\right)  =-i\kappa\nabla^{2}\bar{\psi}\left(  \mathbf{x},t\right)  \;,
\end{equation}
with $\kappa=\frac{1}{2m}$. \ For the real point-split probability density
(just the density matrix in position-position representation),
\begin{equation}
\rho\left(  \mathbf{x},t,\mathbf{a}\right)  \equiv\bar{\psi}\left(
\mathbf{x-a},t\right)  \psi\left(  \mathbf{x+a},t\right)  \;,
\end{equation}
it follows that
\begin{equation}
\partial_{t}\rho\left(  \mathbf{x},t,\mathbf{a}\right)  =\nabla\cdot
\mathbf{J}\left(  \mathbf{x},t,\mathbf{a}\right)  \;.
\end{equation}%
\begin{equation}
\mathbf{J}\left(  \mathbf{x},t,\mathbf{a}\right)  =i\kappa\bar{\psi}\left(
\mathbf{x-a},t\right)  \left(  \nabla_{\mathbf{x}}\psi\left(  \mathbf{x+a}%
,t\right)  \right)  -i\kappa\left(  \nabla_{\mathbf{x}}\bar{\psi}\left(
\mathbf{x-a},t\right)  \right)  \psi\left(  \mathbf{x+a},t\right)  \;.
\end{equation}
However, as above, it is more in line with \cite{CF} to write this current
density as
\begin{equation}
\mathbf{J}\left(  \mathbf{x},t,\mathbf{a}\right)  =i\kappa\nabla_{\mathbf{a}%
}\rho\left(  \mathbf{x},t,\mathbf{a}\right)  \;,
\end{equation}
so that the infinite set of conservation laws become just the second-order
equation
\begin{equation}
\partial_{t}\rho\left(  \mathbf{x},t,\mathbf{a}\right)  =i\kappa\left(
\nabla_{\mathbf{x}}\cdot\nabla_{\mathbf{a}}\right)  \rho\left(  \mathbf{x}%
,t,\mathbf{a}\right)  \;. \label{conservation}%
\end{equation}
We recognize this as just the Fourier transform of the Wigner-Moyal equation
\cite{Wigner,Moyal,ZFC}. \ With
\begin{equation}
f\left(  \mathbf{x},t,\mathbf{p}\right)  =\int d^{n}a\,e^{2i\mathbf{a}%
\cdot\mathbf{p}}\rho\left(  \mathbf{x},t,\mathbf{a}\right)  \;,
\end{equation}
and $H=\kappa p^{2}$, the free-particle Wigner-Moyal equation is given in
terms of Groenewold's $\star$-product \cite{Groenewold} as%
\begin{align}
-i\partial_{t}f\left(  \mathbf{x},t,\mathbf{p}\right)   &  =-2i\kappa
\mathbf{p}\cdot\nabla_{\mathbf{x}}f\left(  \mathbf{x},t,\mathbf{p}\right)
=\left[  H,f\left(  \mathbf{x},t,\mathbf{p}\right)  \right]  _{\star}=H\star
f-f\star H\;,\nonumber\\
H\star f  &  =\kappa\left(  \mathbf{p}-\tfrac{1}{2}i\nabla_{\mathbf{x}%
}\right)  ^{2}f\;,\;\;\;f\star H=\kappa\left(  \mathbf{p}+\tfrac{1}{2}%
i\nabla_{\mathbf{x}}\right)  ^{2}f\;.
\end{align}
Take one $\mathbf{a}$ gradient of (\ref{conservation}) to obtain%
\begin{equation}
\partial_{t}\mathbf{J}\left(  \mathbf{x},t,\mathbf{a}\right)  =i\kappa\left(
\nabla_{\mathbf{x}}\cdot\nabla_{\mathbf{a}}\right)  \mathbf{J}\left(
\mathbf{x},t,\mathbf{a}\right)  \;.
\end{equation}
Or take any number of $\mathbf{a}$ gradients. \ This gives a set of totally
symmetric tensors conserved in precisely the same way.%
\begin{equation}
T_{ij\cdots k}\left(  \mathbf{x},t,\mathbf{a}\right)  \equiv\partial_{a_{i}%
}\partial_{a_{j}}\cdots\partial_{a_{k}}\rho\left(  \mathbf{x},t,\mathbf{a}%
\right)  \;,\;\;\;\partial_{t}T_{ij\cdots k}\left(  \mathbf{x},t,\mathbf{a}%
\right)  =i\kappa\left(  \nabla_{\mathbf{x}}\cdot\nabla_{\mathbf{a}}\right)
T_{ij\cdots k}\left(  \mathbf{x},t,\mathbf{a}\right)  \;.
\end{equation}
Of course, $\mathbf{x}$ derivatives also commute through $\partial_{t}%
-i\kappa\left(  \nabla_{\mathbf{x}}\cdot\nabla_{\mathbf{a}}\right)  $, so
these may also be used to obtain conserved tensors, but as total $x$
derivatives, they are not quite as interesting. \ On the other hand, the
symmetric generators%
\begin{equation}
M_{\left(  ij\right)  }=x_{i}\partial_{x_{j}}+x_{j}\partial_{x_{i}}%
-a_{i}\partial_{a_{j}}-a_{j}\partial_{a_{i}}%
\end{equation}
also commute with $\nabla_{\mathbf{x}}\cdot\nabla_{\mathbf{a}}$, and may be
used to construct other conserved densities.

It is amusing that the conservation laws (\ref{conservation}) are
\textit{exactly} of the form as in the introductory remarks of \cite{CF}, with
complexification of the second-derivatives' coefficient. \ Unfortunately, if
that coefficient is not purely imaginary, then the construction falters for
real point-splittings. \ In other words, for real functions obeying the real
diffusion equation this particular collection of conservation laws does not
exist. \ Still, it is also amusing that one could indeed think of the
point-splitting as introducing extra dimensions, and doing so by
dimension-doubling. \ The wave functions involved in the currents are then
just two distinct types of particular free wave functions in twice the number
of spatial dimensions, namely $\psi\left(  \mathbf{x},t,\mathbf{a}\right)
=\psi\left(  \mathbf{x+a},t\right)  $ and $\bar{\psi}\left(  \mathbf{x}%
,t,\mathbf{a}\right)  =\bar{\psi}\left(  \mathbf{x-a},t\right)  $. \ They each
obey the extra-dimensional equations
\begin{equation}
\partial_{t}\psi\left(  \mathbf{x},t,\mathbf{a}\right)  =i\frac{\kappa}%
{2}\left(  \nabla_{\mathbf{x}}^{2}+\nabla_{\mathbf{a}}^{2}\right)  \psi\left(
\mathbf{x},t,\mathbf{a}\right)  \;,\;\;\;\partial_{t}\bar{\psi}\left(
\mathbf{x},t,\mathbf{a}\right)  =-i\frac{\kappa}{2}\left(  \nabla_{\mathbf{x}%
}^{2}+\nabla_{\mathbf{a}}^{2}\right)  \bar{\psi}\left(  \mathbf{x}%
,t,\mathbf{a}\right)  \;.
\end{equation}
There are other solutions of these free particle equations, of course, but as
in \cite{CF} we need only make use of particular solutions to encode the
conservation laws.

\section{Generalized Bateman equation}

In previous investigations \cite{Fairlie}, the Euler-Monge equation was used
to obtain the prototypical Bateman equation \cite{Bateman}. \ This section
addresses the question of what happens when the pressure term is added, and
gives solutions of the resulting modified Bateman equation.

First of all we notice that the equation
\begin{equation}
\frac{\partial^{2}u}{\partial t^{2}}=\frac{\partial}{\partial x}\left(
u\frac{\partial u}{\partial t}\right)
\end{equation}
holds whatever the pressure term is, provided it is dependent only upon $x$.
\ This implies the existence of a scalar $\phi(x,t)$ such that
\begin{equation}
\frac{\partial u}{\partial t}=\frac{\partial\phi}{\partial x}%
\;,\;\;\;u\frac{\partial u}{\partial t}=\frac{\partial\phi}{\partial t}\;.
\label{PhiEqns}%
\end{equation}
That is,%
\begin{equation}
u=\frac{\partial\phi/\partial t}{\partial\phi/\partial x}\;. \label{uAsPhis}%
\end{equation}
This and (\ref{pressure}) imply that $\phi$ satisfies a Bateman-like equation,
in interaction with an external field:
\begin{equation}
(\phi_{x})^{2}\phi_{tt}-2\phi_{x}\phi_{t}\phi_{tx}+(\phi_{t})^{2}\phi
_{xx}=(\phi_{x})^{3}g(x)\;. \label{Bateman-like}%
\end{equation}
Note that the RHS preserves homogeneity in $\phi$, but breaks homogeneity in
$t$ and $x$, in general. \ In the particular case where $g(x)=k$, a constant,
and $u=\frac{1}{t}\left(  -x+\frac{kt^{2}}{2}\right)  $, $\phi$ is determined
by (\ref{PhiEqns}) to be $\phi=\frac{1}{2t^{2}}\left(  x+\frac{kt^{2}}%
{2}\right)  ^{2}$. \ Since any function of $\phi$ will also satisfy the
Bateman-like equation (as is clear from (\ref{uAsPhis})), we might as well
take the square root of this for verification of (\ref{Bateman-like}).

If $F(\phi),\ G(\phi)$ are arbitrary differentiable functions of $\phi$, then
the general solution of the Bateman-like equation when the pressure-gradient
is constant is given by solving implicitly for $\phi$ the following equation:
\begin{equation}
(x+\frac{kt^{2}}{2})\,F(\phi)+t\,G(\phi)=\mathrm{constant\;.}
\label{ImplicitBateman}%
\end{equation}
Direct calculation establishes this result. \ We find%
\begin{equation}
\phi_{x}=\frac{-F}{(x+\frac{kt^{2}}{2})F^{\prime}+tG^{\prime}}\;,\;\;\;\phi
_{t}=\frac{ktF+G}{F}\phi_{x}\;,
\end{equation}
and hence%
\begin{align}
\phi_{x}\phi_{xt}  &  =\phi_{t}\phi_{xx}+\frac{FG^{\prime}-GF^{\prime}}{F^{2}%
}\phi_{x}^{3}\;,\nonumber\\
\phi_{t}\phi_{xt}  &  =\phi_{x}\phi_{tt}-\frac{FG^{\prime}-GF^{\prime}}{F^{2}%
}\phi_{x}^{2}\phi_{t}-k\phi_{x}^{2}\;.
\end{align}
Substitution of these last two expressions into the Bateman-like equation
(\ref{Bateman-like}) proves the claim. \ The implicit equation
(\ref{ImplicitBateman}) gives the general solution to (\ref{Bateman-like}%
),\ for constant pressure-gradient, as it depends upon two independent
arbitrary functions. \ 

A similar result can be found in the case of a linear pressure-gradient term,
$g(x)=k^{2}x$, which is the unique example that preserves homogeneity in $x$
for (\ref{Bateman-like}). \ Here, the general solution is given by the
implicit solution for $\phi$ of the following equation:
\begin{equation}
\frac{\sin kt}{x}\,F(\phi)+\frac{\cos kt}{x}\,G(\phi)=\mathrm{constant\;.}%
\end{equation}
Once again, direct calculation establishes this result. \ This is the second
known case where a generalized form of the Bateman equation is integrable, the
other being the so-called ``two dimensional Born-Infeld equation''
\cite{Nutku,Mulvey}.

More general pressure-gradients appear to be less tractable for the
Bateman-like equation, even implicitly, without resorting to the Euler-Monge
form of the equation. \ Note that the recipe which worked to obtain implicit
solutions of the driven Euler-Monge equation for arbitrary $g\left(  x\right)
$ is not as useful for (\ref{Bateman-like}), since thinking of $t\left(
x,\phi\right)  $ instead of $\phi\left(  x,t\right)  $ does not linearize that equation.

\section{Summary}

We have considered here the effects of given pressure-gradients on the
one-dimensional flows of the Euler-Monge equation, and on the corresponding
driven diffusion/Schr\"{o}dinger equations expressed in an extra dimension.
\ The extra dimension method provides some additional insight and
computational tools beyond those given by the method of characteristics and
the technique of constructing implicit solutions, which we have discussed and
compared. \ While we have touched on some elementary features of conservation
laws for the free particle Schr\"{o}dinger case, including their relation to
the Wigner-Moyal equation, a full analysis of higher-dimensional driven flows
along the lines of \cite{CF}\ remains to be given \cite{note5}. \ The effects
of viscosity have also not been considered. \ (For a recent study including
viscosity effects, see \cite{Boldyrev}.) \ We hope to return to these issues
in a subsequent paper.

\paragraph{Acknowledgements:}

We thank the referee for raising the issue of shock fronts, and J Nearing for
bringing to our attention P Lax's concise monograph on the subject. \ This
research was supported in part by a Leverhulme Emeritus Fellowship and by NSF
Award 0073390.


\begin{thebibliography}{9}
\bibitem {note1}If $(x,a)$ are conjugate complex co-ordinates, $x=\frac{1}%
{2}(x_{1}+ix_{2}),\ a=\frac{1}{2}(x_{1}-ix_{2}),$ then recall that
$\frac{\partial^{2}}{\partial x\partial a}=\frac{\partial^{2}}{\partial
x_{1}^{2}}+\frac{\partial^{2}}{\partial x_{2}^{2}}$.
\bibitem {note2}In this regard, note in (\ref{LinearSeriesSolution}) that
$\left(  \sin kt\right)  ^{j}\frac{d^{j}}{dx^{j}}=\left(  \tan kt\right)
^{j}\frac{d^{j}}{d\left(  x/\cos kt\right)  ^{j}}$. \
\bibitem {note3}This transformation is analogous to the relation between free
point particle motion, with $x_{\text{free}}\left(  t\right)  =x_{0}+v_{0}t$,
and simple harmonic motion, with $x_{\text{SHO}}\left(  t\right)  =x_{0}\cos
kt+\frac{1}{k}v_{0}\sin kt$, whose two forms are connected by a dynamical
rescaling and substitution for the time variable: $x_{\text{SHO}}\left(
t\right)  /\cos kt=x_{\text{free}}\left(  t\rightarrow\frac{1}{k}\tan
kt\right)  .$
\bibitem {note4}As further illustration of this transformation technique, if
$u(x,t)$ is a solution of the undriven equation (\ref{EulerMonge}), then
$\mathfrak{u}\left(  x,t\right)  \equiv u\left(  x+s\left(  t\right)
,t\right)  +ds\left(  t\right)  /dt$ satisfies $\partial_{t}\mathfrak{u}%
=\mathfrak{u}\partial_{x}\mathfrak{u}+d^{2}s\left(  t\right)  /dt^{2}.$
\bibitem {note5}However, it should be readily apparent that the
higher-dimensional flows produced by the special cases of constant or linear
pressure-gradients can be expressed in terms of undriven solutions, with the
same interpretations and forms as given above following (\ref{USolution}) and
(\ref{LinearSeriesSolution}).
\end{thebibliography}
\end{document}